# Probing and tuning inelastic phonon conductance across finite-thickness interface

Takuru Murakami[1], Takuma Hori[1], Takuma Shiga[1], and Junichiro Shiomi[1,2,*]

[1]*Department of Mechanical Engineering, The University of Tokyo, Bunkyo, Tokyo, 113-8656, Japan*

[2]*Japan Science and Technology Agency, PRESTO, 4-1-8, Kawaguchi, Saitama, 332-0012, Japan*

E-mail: shiomi@photon.t.u-tokyo.ac.jp

Phonon transmission across an interface between dissimilar crystalline solids is calculated using molecular dynamics simulations with interatomic force constants obtained from first principles. The results reveal that although inelastic phonon-transmission right at the geometrical interface can become far greater than the elastic one, its contribution to thermal boundary conductance (TBC) is severely limited by the transition regions, where local phonon states at the interface recover the bulk state over a finite thickness. This suggests TBC can be increased by enhancing phonon equilibration in the transition region for instance by phonon scattering, which is demonstrated by increasing the lattice anharmonicity.

Heat conduction through the boundary between dissimilar crystalline solids is a fundamental property that determines thermal conductivity of materials or system with contacts, grains, and layers. The conductance through the interface is usually quantified as thermal boundary conductance (TBC), a reciprocal of thermal boundary resistance. Understanding TBC has become more and more important particularly in predicting and designing thermal performance of nanostructured materials such as thin films, superlattices, and nanograins. The overall thermal conductivity of such materials with scales smaller than phonon mean free paths is determined by TBC in addition to the size effect of intrinsic thermal conductivity.

A widely accepted physical picture describes TBC in terms of the mismatch of bulk phonon properties. The simplest description is provided by the acoustic mismatch model[1] and diffuse mismatch model[2], which have been commonly used to provide reference TBC values for perfect interface at low temperature and rough interface at high temperature, respectively. More advanced models have also been proposed to further account for full phonon dispersions[3], electron-phonon coupling[4], and multiple phonon scattering.[5] These models assume phonon transmission across the interface to be elastic, coinciding with calculations performed for single-mode transmission under groundstate (zero Kelvin), such as wavepacket[6], linear lattice dynamics[7, 8], and nonequilibrium Green's function[9] methods.

On the other hand, inelastic phonon transmission has been suggested to contribute significantly to TBC at high temperature[10], which is supported by the linear increase of TBC with temperature observed in molecular dynamics (MD) simulations[11-13] and experiments.[14, 15] Apart from some phenomenological models proposed to accounting for inelastic transmission[16-18], it is only recently that the phonon transmission including inelastic contribution can be directly calculated using the phase-space trajectories in MD simulations.[19] For a well-defined weakly-bonded interface between CNT and silicon, inelastic phonon transmission extending beyond the silicon cut-off frequency has been found.

The result is consistent with the contribution of inelastic energy transport from CNT to surrounding matrix found by transient MD simulations.[20]

While the inelastic phonon transmission at the geometrical interface may well represent the picture of TBC for a well-defined interface, the relation between the transmission function and the TBC is not clear when the interface has a finite characteristic thickness. It is well known that even perfectly coherent interfaces can give rise to local phonon states that extents some distance away from the interface. This can be also related to the nonlinear temperature gradient regime around the interface often observed in steady nonequilibrium MD (NEMD) simulations.[21] We note that such finite-thickness transition region (TR) is omnipresent particularly among interfaces with the elastic properties similar to those of the materials on both sides, and should be as general as sharp interfaces if not more.

In this work, we clarify how inelastic phonon transmission does or does not contribute to TBC across the finite-thickness interface. We calculate the local phonon states and inelastic phonon transmission by performing equilibrium MD (EMD) using the interatomic force constants (IFCs) obtained from first principles. We investigate two representative thermal interfaces, a sharp interface between PbTe/PbS and a finite-thickness interface between Si/Ge that exhibit distinct difference in the role of inelastic phonon transmission. A comparative study with steady NEMD simulations identify that the mode-dependent TR between the bulk and interface needs to be considered to understand the full picture of inelastic phonon transmission. Understanding of the mechanism further leads us to evaluate controllability of TBC by tuning material properties.

The anharmonic IFCs were expressed in the form of Taylor expansion of the force respect to the atomic displacement $u$ around the equilibrium positions,

$$-F_i^\alpha = \Pi_i^\alpha + \sum_{j,\beta} \Phi_{ij}^{\alpha\beta} u_j^\beta + \frac{1}{2!} \sum_{jk,\beta k} \Psi_{ijk}^{\alpha\beta\gamma} u_j^\beta u_k^\gamma + \frac{1}{3!} \sum_{jkl,\beta kl} X_{ijkl}^{\alpha\beta\gamma\delta} u_j^\beta u_k^\gamma u_l^\delta , \tag{1}$$

where, $\Phi$, $\Psi$, and $X$ are harmonic, cubic, and quartic IFCs, respectively. The indices $i$, $j$, $k$,

and $l$ are atom indices, and $\alpha$, $\beta$, $\gamma$, and $\delta$ represent Cartesian components. IFCs of PbTe, PbS, and Si single crystals are obtained by the real-space displacement method[22] based on random displacements of atoms[23]. The lattice constants of PbTe, PbS, and Si were set to 6.482Å, 5.962Å, and 5.411Å, which give minimum total ground-state energies. Validity of the methodology and has been shown for various materials[22-25].

At the PbTe/PbS interface, the IFCs between PbTe and PbS were described using the local virtual crystal approximation[23].

$$\Phi_{ij}^{\text{PbTe/PbS}} = \left(1 - \xi_{ij}\right)\Phi_{ij}^{\text{PbTe}} + \xi_{ij}\Phi_{ij}^{\text{PbS}} \tag{2}$$

Here, $\xi_{ij}$ is a scalar that weights how much the surrounding first neighbor atoms to the $i$th atom resemble those of pure PbS (See Ref. 23 for details). The cubic and quartic IFCs can also be treated in the pairwise fashion since only the nearest neighbors are considered for the anharmonic IFCs. This local virtual crystal approximation has successfully reproduced substitution alloy effect of lead chalcogenides[23]. For the Si/Ge system, we used only the IFCs of Si crystal because the IFCs of Si and Ge crystals were found to be transferable. The transferability has been known for harmonic terms[26] but we in addition found that this is true also for the cubic terms (See supplementary information for details).

A typical EMD simulation was performed for PbTe/PbS and Si/Ge systems in fully periodic cells with sizes of 124.4Å×31.11Å×31.11Å and 216.3Å×27.04Å×27.04Å, respectively. In an NEMD simulation, the end layers were fixed and the two layers adjacent to them were steadily heated and cooled, and periodic boundary conditions were applied in the cross-sectional directions. In both EMD and NEMD simulations, the size of the system is made large enough to prevent the size effect in the phonon transmission function[21].

Figures 1(a) and (b) show temperature profiles of PbTe/PbS and Si/Ge system obtained by NEMD. TBC can be calculated by dividing the temperature jump at the interface with the heat flux through the system. Here the temperature jump is identified by extrapolating the linear temperature profile in the bulk region. The bulk temperature gradient was consistent with the bulk thermal conductivity. At the PbTe/PbS interface, temperature discontinuously jumps at the interface, while, at the Si/Ge interface, temperature transits continuously over several layers adjacent to the geometrical interface. This indicates that the phonon transmission process through the interface is different between the two systems. In the case of PbTe/PbS [Fig.2(a)], we can simply describe the system as the interface S directly sandwiched by the two mediums that effectively has bulk phonon properties. On the other hand, in case of Si/Ge [Fig.2(b)], TR extends over a finite thickness around S, which is denoted as α and β.

Now we calculate the phonon transmission function at S for PbTe/PbS and Si/Ge systems. Based on the partial derivative of Hamiltonian of the sub-cells divided by the interface, heat flux between the left (A) and right (B) sub-cells can be expressed in terms of the cross-correlation of the atomistic velocities ($v$) and forces on B from A ($F_j^{A \to B}$) as $J_{A \to B} = \lim_{\tau \to 0} \sum_{j \in B} \left\langle F_j^{A \to B}(t+\tau) \cdot v_j(t) \right\rangle / a$ , where $a$ is the cross-sectional area. The mode-dependent heat flux can then be obtained by Fourier transformation, and the phonon transmission function can be derived by comparing the Landauer and atomistic expression of heat flux[19].

$$\Theta(\omega) = \frac{1}{2ak_B T} \left| \mathrm{Re} \left[ i\omega \sum_{\alpha,\beta} \sum_{i \in A, j \in B} \Phi_{ij}^{\alpha\beta} \left\{ u_i^\beta(\omega) - u_j^\beta(\omega) \right\} u_j^{\alpha*}(\omega) \right] \right|, \qquad (3)$$

where $\omega$ and $k_B$ are the phonon frequency and Boltzmann constant. Note that anharmonic effects are fully incorporated through the anharmonic IFCs in Eq.(1). The transmission here is essentially a frequency dependent thermal conductance, and thus, TBC can be calculated by integrating the transmission function over frequency.

$$K = k_B \int_0^{\omega_{max}} \Theta(\omega) d\omega. \qquad (4)$$

Note that the sign of $J_{A \to B}$ can also be negative and it becomes zero when averaged in time since the simulation is at equilibrium. Therefore, as an empirical treatment, we take absolute value of the calculated quantity in Eq.(3) with an intention to consider phonon transmission in the same direction with $J_{A \to B}$.

Figures 3(a) and (b) show transmission function of PbTe/PbS and Si/Ge interfaces at 300K, respectively. The frequency domain can be divided into two regimes: (I) $0 < \omega < \omega_{max}^{L}$ and (II) $\omega_{max}^{L} < \omega < \omega_{max}^{H}$, where L and H are the materials with lower and higher phonon frequencies. In current cases, L is PbTe or Ge, and H is PbS or Si. The transmission function in regime (I) is attributed to elastic phonon scattering, i.e. the phonons incoming to and outgoing from the interface has the same frequency ("elastic channel"). On the other hand, in regime (II), inelastic scattering contributes to the transmission ("inelastic channel"). Note that the inelastic process can also contribute to the transmission function in regime (I) but nevertheless the transmission function in regime (II) is a good indication of the contribution of the inelastic channel.

The comparison of PbTe/PbS and Si/Ge reveals a case-dependence of the extent of the inelastic channel. In case of PbTe/PbS, the inelastic channel is moderate compared with the elastic one, and contributes to only about 10% of the total TBC. In contrast, in case of Si/Ge, the inelastic channel contributes to 70% of the total TBC. We first guessed this to be due to the difference in lattice anharmonicity, which is expected to widen the inelastic channel. This was checked by varying anharmonicity at S by turning on and off the anharmonic terms in Eq.(1), however, it had negligible influence on the transmission function, which is consistent with a recent observation by NEMD[27]. We then turned to the local phonon density of states (DOS), which was investigated for each layer by calculating the power spectra of atom velocities. As shown in Fig.4(a), the DOS on the Ge-side of S (0th layer) exhibits a peak that extends beyond the cutoff frequency of bulk Ge. Similar peak is observed on the Si-side and agrees with the peak position in the phonon transmission function [Fig.3(b)], suggesting that the inelastic channel is associated with the local phonon states at S.

In case of PbTe/PbS, the obtained TBC calculated from EMD ($0.33\text{GWm}^{-2}\text{K}^{-1}$) and NEMD ($0.34\text{GWm}^{-2}\text{K}^{-1}$) was in excellent agreement. Contrastingly, in case of Si/Ge, TBC calculated by EMD ($2.8\text{GWm}^{-2}\text{K}^{-1}$) and NEMD ($0.27\text{GWm}^{-2}\text{K}^{-1}$) resulted in an order of magnitude difference. TBC obtained by our NEMD is in good agreement with the values obtained by lattice dynamics ($0.28\text{GWm}^{-2}\text{K}^{-1}$)[8] and NEMD ($0.32\text{GWm}^{-2}\text{K}^{-1}$)[11] using the empirical Stillinger Weber potential, and recent NEGF calculation using first-principles-based IFCs similar to this work ($0.28\text{GWm}^{-2}\text{K}^{-1}$)[28].

The difference between TBCs of EMD and NEMD for Si/Ge should be due to TR, which is included in NEMD but not in EMD. Now we postulate the picture of TBC in Fig.2(b) introducing the additional vertical interfaces A'(B') between bulk and α(β). In this picture, the process of interfacial thermal transport from Si to Ge takes place as: phonons transmit (i) from bulk Si to α according to the transmission function at A', (ii) from α to β according to its transmission function at S, and (iii) from β to bulk Ge according to the transmission function at B'. Note that in the mind of Landauer theory, we ignore phonon scatterings inside α and β assuming ballistic transport.

Now the question is how to define the thickness of α and β. In this work, we take the decay-length of the correlation coefficient between the layer at S and layers along to the direction of heat flux. Similarly to the transmission function, the thickness of TR is expected to depend on phonon modes, which is here evaluated as a frequency-dependent property. The frequency dependent correlation coefficient $h(l,\omega)$ is calculated as

$$h(l,\omega) = \frac{C_{0l}(\omega) \cdot C_{0l}^{*}(\omega)}{C_{00}(\omega) \cdot C_{00}(\omega)}, \qquad (4)$$

where $C$ is the cross spectrum defined as

$$C_{mn}(\omega) = \sum_{i \in m} \sum_{j \in n} \mathbf{v}_i^{*}(\omega) \cdot \mathbf{v}_j(\omega). \qquad (5)$$

Here, $m$ and $n$ are the layer indices $(0,1,\cdots)$ counted from the layer at S. The thickness of TR, $L(\omega)$, is then obtained by fitting an exponential function, $h(q,\omega) = \exp(-l/L(\omega))$. The calculated $L(\omega)$ on the Si and Ge sides [Fig.4(b)] ranges from 2 to 1000 layers depending on the phonon frequency. This means that the surface states are localized near S for high-frequency phonons and extend for a significantly long distance for low-frequency ones. Note the results did not change with the system size.

We then calculate the total phonon transmission function of the entire TR consisting of series of TBC at A', S and B'. The inverse of the total phonon transmission function is calculated as the series sum:

$$\frac{1}{\Theta_{\text{total}}(\omega)} = \frac{1}{\Theta_{A'}(\omega, L_{\text{Si}}(\omega))} + \frac{1}{\Theta_{S}(\omega)} + \frac{1}{\Theta_{B'}(\omega, L_{\text{Ge}}(\omega))}. \qquad (6)$$

One technical problem here is that $L$ of some low-frequency phonons exceed half the length of the simulation cell (40 layers). This was remedied by cutting off $L$ of the low frequency phonons at 40 layers since their contribution to TBC is minor compared with the rest as can be seen in the transmission function [Fig.3(b)]. This was confirmed by varying the length of the system, which made negligible difference in TBC. This also means that the effective thermal interface is still a local property near S and justifies the linear extrapolation done in NEMD on extracting TBC.

The obtained total phonon transmission function including TR is shown in Fig.4(c). Furthermore, TBC calculated from the transmission function is plotted together with that obtained by NEMD for different temperatures in Fig.4(d). Figure 4(d) clearly shows that by taking the total transmission into account, TBC of EMD agrees with that of NEMD. Firstly, this suggests the validity of the series resistance model in Fig.2. Secondly, this confirms that an order of magnitude difference between Si/Ge-TBCs calculated by EMD ($2.8 \text{GWm}^{-2}\text{K}^{-1}$) and NEMD ($0.27 \text{GWm}^{-2}\text{K}^{-1}$) is due to the missing contribution from TR. On the other hand, for PbTe/PbS, $L$ is short enough so that the phonon transmission at S describes TBC well.

It is interesting to note that obtained TBC at S ($2.8 GWm^{-2}K^{-1}$) is in reasonable agreement with the thermal conductance of Si/Ge superlattice:~$2 GWm^{-2}K^{-1}$ per layer (i.e. per interface) measured in experiments[29, 30]. This makes sense if we think that the obtained thermal conductance per layer is similar to TBC at S because the superlattice period thickness (3nm and 4.4nm in Ref.29 and Ref.30, respectively) is smaller than $L$. This may explain the reported discrepancy between the calculated TBC of Si/Ge interface and those extracted from the superlattice experiments[28].

The total phonon transmission function of Si/Ge interface in Fig.4(c) shows that despite the dominant contribution from the inelastic channel to the phonon transmission across S [Fig.3(b)], the overall TBC is still dominated by the elastic channel. This is because the thermal resistance between the bulk and TR (A' and B') screens the high frequency phonons before reaching S, even though the inelastic channel at S has a potential to transmit high frequency phonons. This also means that the phonon distribution in the TR is strongly nonequilibrium and skewed towards low frequency otherwise the inelastic channel will be utilized.

The above discussion suggests, if we are to promote heat conduction across the interface, one way to make a better use to the inelastic channel is to equilibrate the phonon distribution in TR, for instance, by enhancing phonon-phonon scattering in the boundary regime. Here, we have demonstrated it by increasing the anharmonicity in TR by scaling the anharmonic IFCs with a constant value $\gamma$ (=1.0,1.1,1.2, and 1.5). This approach is convenient for proof of principles since the eigenstates do not change. As shown in Fig.5, by increasing $\gamma$ significantly reduces the temperature drop at the interface as expected. As a result, TBC significantly and monotonically increases (0.27,0.31,0.34, and $0.37 GWm^{-2}K^{-1}$, respectively) with $\gamma$. This clearly demonstrates the controllability of TBC by utilizing the inelastic channel.

In summary, we calculated frequency-dependent phonon transmission function of PbTe/PbS and Si/Ge interfaces by EMD with first-principles-based IFCs. The comparative study of EMD and NEMD suggests that TBC through the Si/Ge interface with finite-thickness TR can be well described by series model of thermal resistance between the bulk and TR and that right at the interface. Although the phonon transmission function at the interface exhibit potential to transport heat through the inelastic channel, the total transmission is bottlenecked by the transmission between the bulk and transient region. The inelastic channel can be more effectively utilized by enhancing phonon equilibration in TR for instance through phonon-phonon scattering, which is demonstrated by increasing the anharmonicity in TR. The current work suggests possibility to control TBC by utilizing the inelastic channel.

## Acknowledgments

This work is partially supported by JST, PRESTO and JSPS, KAKENHI 26709009.

**Figure Captions**

**Fig.1.** The temperature profiles of (a) PbTe/PbS and (b) Si/Ge systems in steady state. The red and blue triangles indicate the thermostated layers at hot and cold temperatures. The orange and green dash lines show the fitting of the temperature profiles in the bulk regions.

**Fig. 2.** The schematic images of (a) PbTe/PbS and (b) Si/Ge systems. The domains named α and β in Si/Ge system indicate the transition region. A' (B') is the boundary between bulk Si (Ge) and domain α (β), and $\Theta_{A'}$ ($\Theta_{B'}$) is the corresponding phonon transmission function.

**Fig. 3.** The phonon transmission functions at the geometrical interface S at (a) PbTe/PbS and (b) Si/Ge interfaces. The dash lines are the maximum frequencies of the bulk materials.

**Fig. 4.** (a) Local phonon density of states of each layer counted from the interface S, (b) the thickness of the transition region on Si (red solid line) and Ge (blue dashed line) side, (c) the overall phonon transmission of Si/Ge including the transition regions, and (d) comparison of the thermal conductance with that obtained by NEMD. The green dashed lines denote half the length of the simulation cell (40 layers).

**Fig. 5.** The temperature profile of the Si/Ge system in which the anharmonicity of boundary domain α is enhanced by scaling anharmonic IFCs. The black, orange and red lines indicate the result of scaling factor 1.0, 1.1 and 1.5.

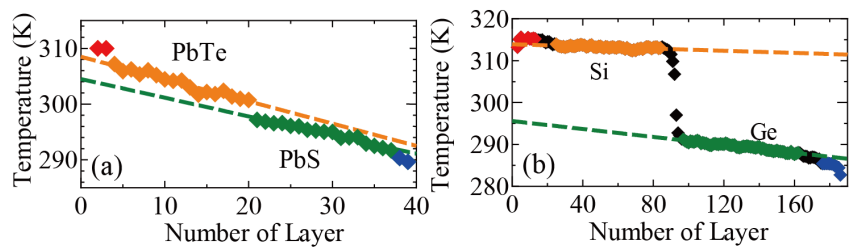

Fig. 1

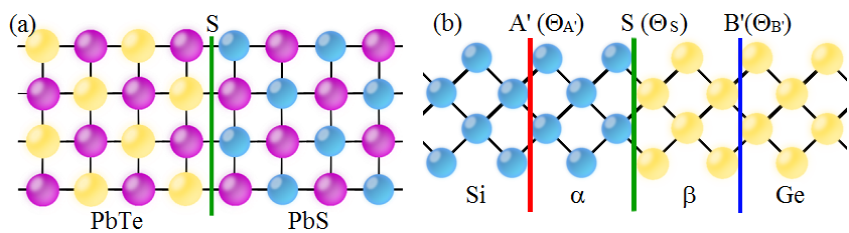

Fig. 2

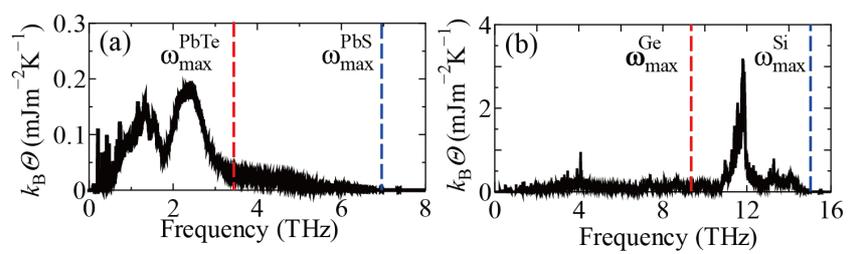

Fig. 3

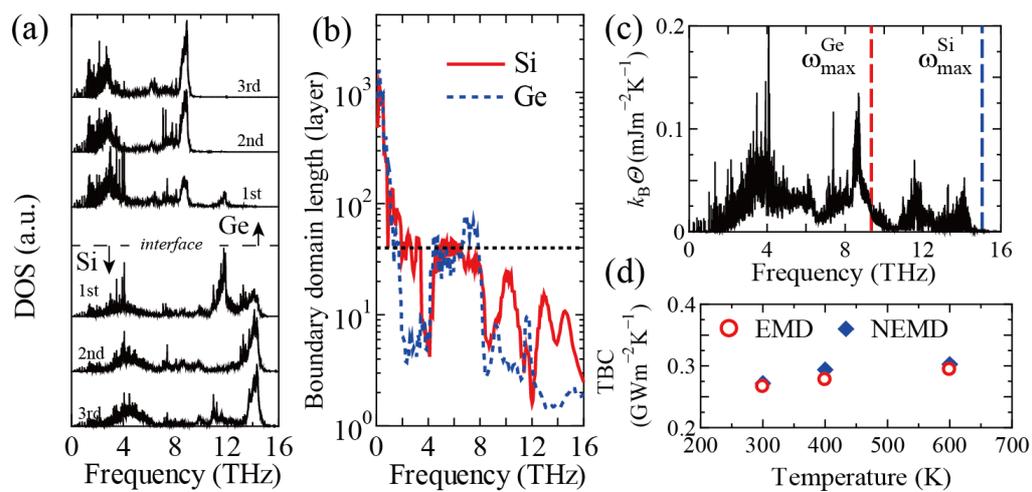

Fig. 4

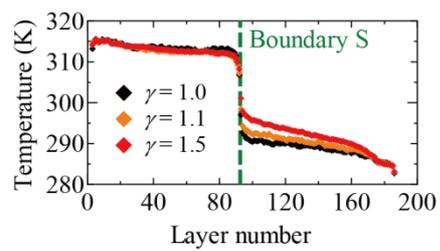

Fig. 5